\documentclass[twocolumn,showpacs,showkeys,preprintnumbers,amsmath,amssymb,aps]{revtex4-1}

\usepackage{graphicx}
\usepackage{dcolumn}
\usepackage{bm}

\begin{document}

\title{Impact of heavy hole-light hole coupling on optical selection rules in GaAs quantum dots}

\author{T. Belhadj$^1$}
\author{T. Amand$^1$}
\author{A. Kunold$^{1,2}$}
\author{C.-M. Simon$^{1,3}$}
\author{T. Kuroda$^4$}
\author{M. Abbarchi$^4$}
\author{T. Mano$^4$}
\author{K. Sakoda$^4$}
\author{S. Kunz$^1$}
\author{X. Marie$^1$}
\author{B. Urbaszek$^1$}
\affiliation{%
$^1$Universit\'e de Toulouse, INSA-CNRS-UPS, LPCNO, 135 Av. Rangueil, 31077 Toulouse, France}
\affiliation{%
$^2$Departamento de Ciencias B\'asicas, UAM-A, Col. Reynosa Tamaulipas, 02200 M\'exico D.F., M\'exico}
\affiliation{%
$^3$Universit\'e de Toulouse, CNRS-UPS, LCAR, IRSAMC, 31062 Toulouse, France}
\affiliation{%
$^4$National Institute for Material Science, Namiki 1-1, Tsukuba 305-0044, Japan}

\date{\today}

\begin{abstract}
We report strong heavy hole-light mixing in GaAs quantum dots grown by
droplet epitaxy. Using the neutral and charged exciton emission as a
monitor we observe the direct consequence of quantum dot symmetry
reduction in this strain free system.
By fitting the polar diagram of the emission with simple
analytical expressions obtained from k$\cdot$p theory
we are able to extract the mixing that arises from
the heavy-light hole coupling due to the geometrical
asymmetry of the quantum dot.
\end{abstract}

\maketitle

\begin{figure}
\includegraphics[width=0.48\textwidth]{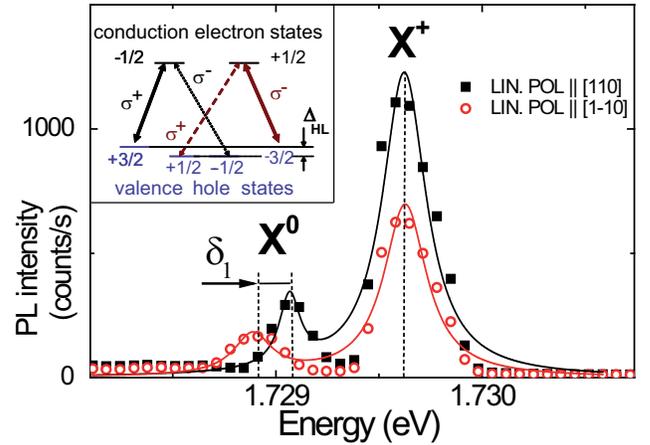}
\caption{\label{fig:fig1} (Color online) PL spectra of an individual GaAs quantum dot. Emission from the neutral exciton X$^0$ with fine structure splitting $\delta_1=170\mu$eV and the positively charged exciton X$^+$. Full squares (open circles) correspond to linearly polarized emission along $[110]$ ($[1\bar{1}0]$). Solid lines indicate fits obtained with Lorentzian line shapes. The selection rules for optical transition between conduction electrons and valence holes are
shown in the inset.
}
\end{figure}

A large variety of promising applications for self assembled semiconductor quantum dots in photonics and spin electronics are based on the discreteness of the interband transitions \cite{Michler2000} and on long carrier spin relaxation times \cite{Jonker08,lombez2007}. Combining the two characteristics allows electrical tuning of the light polarization of single photon emitters based on quantum dots \cite{Strauf07}. Optical preparation, manipulation and read-out of a single spin state \cite{Henne2009,berezovsky08,press08,xu09} are governed by the optical selection rules \cite{lampel68,md2008}. 
For excitons based on pure angular momentum eigenstates for electrons ($m_s=\pm 1/2$) and for heavy holes ($m_j=\pm 3/2$), a given light polarization can be associated unambiguously with the optical creation of a certain spin state (solid arrows in the inset of Fig. \ref{fig:fig1}). In the common case of [100] growth (defining the quantization axis $z$) the symmetry of the bulk semiconductor implies that the $x$ and $y$ axis, [110] and $[1\bar{1}0]$,  are equivalent ($D_{2d}$). The conduction electron is well described as an isotropic particle of spin $\pm 1/2$. In contrast, valence hole states are non-isotropic as the x-y equivalence is lifted in realistic dots due to strain and/or shape and in plane anisotropy \cite{koudinov04}. Despite being separated in energy by tens of meV ($\Delta_{HL}$ in the inset of Fig. \ref{fig:fig1}), substantial mixing between valence heavy hole and light hole (HH-LH) states has been reported in strained II-VI and III-V quantum dots \cite{koudinov04,leger07,Krizhanovskii,eble09,kurtze09}. The resulting non pure selection rules in combination with long carrier relaxation times have allowed the implementation of very original, high fidelity ($\ge 99\%$) electron and hole spin initialization schemes \cite{atature06,gerardot08}. In addition, HH-LH mixing has been identified as the key parameter at the origin of the dipole-dipole nuclear spin mediated hole spin dephasing in quantum dots \cite{eble09,Steel09}. The dominant physical origin of HH-LH mixing has been attributed to lattice strain \cite{koudinov04,leger07,kowalik07}.

This was our motivation to study \textit{strain free}, individual GaAs/AlGaAs quantum dots by molecular droplet epitaxy \cite{koguchi91,abbarchi08,belhadj08,keizer10}, see \cite{plumhof10} for alternative
growth method.
We performed detailed, angle dependent polarization analysis of the photoluminescence (PL) emitted parallel to the quantum dot growth direction. We observe, even in the absence of strain, strong evidence for HH-LH mixing revealed in the emission of the positively charged exciton X$^+$ (trion: 2 valence holes + 1 conduction electron) and the neutral exciton X$^0$ (1 hole + 1 electron). Our experiments show that (i) an asymmetry in the confinement potential due to quantum dot elongation (ii) the main axis of the quantum dot potential not coinciding with the crystallographic axis $[110]$ and $[1\bar{1}0]$ contribute to HH-LH mixing. These interpretations are confirmed by fitting the data with a $6$ band k$\cdot$p model.

Droplet epitaxy allows quantum dot self assembly in lattice matched systems as it is not strain driven. The sample investigated here consists of: n-type GaAs (100) substrate, n-GaAs buffer (400 nm), GaAs (50 nm), Al$_{0.3}$Ga$_{0.7}$As (100 nm), a single layer of GaAs QDs on a Wetting Layer, Al$_{0.3}$Ga$_{0.7}$As (100 nm), GaAs (10 nm).
 Detailed spectroscopic analysis, carried out with a confocal microscope as in \cite{belhadj08}, shows (residual) p-type doping leading to the observation of both X$^0$ and X$^+$. The PL light emitted along the growth axis $z$ passes through an achromatic half-wave plate, which can be rotated in the $xy$ plane, in front of a linear Glan-Taylor polarizer before detection. The cw He-Ne laser with an energy of $1.95$eV excites carriers non-resonantly in the AlGaAs barrier. The laser is linearly polarized to avoid any dynamic nuclear polarization and to assure that an equal proportion of carriers in spin up and down states relaxes towards the dot. All the experiments are performed at $T=4$K.

\begin{figure}
\includegraphics[width=0.48\textwidth]{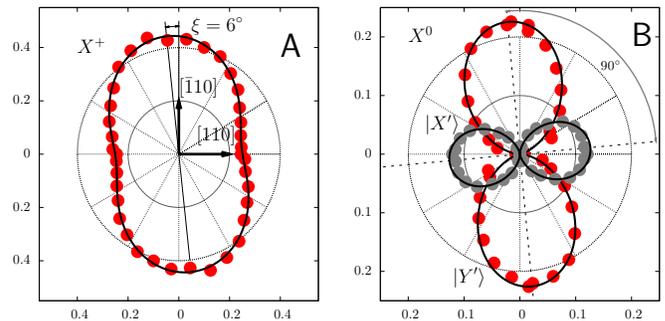}
\caption{\label{fig:fig2} (Color online)
(A) Polar plot of the charged exciton X$^+$ emission intensity for different
positions of the linear polarization analyzer relative to $[110]$
crystallographic direction. (B) Neutral excitons $X'$ (gray) and $Y'$ (red).
The solid (black) lines indicate the theoretical calculation.
by fitting with Eqs. (\ref{noprime})-(\ref{yprime}) we obtain
$\beta=0.25$.
}
\end{figure}

Single dot PL spectra show three transitions that we attribute to the X$^0$, the X$^+$ and the neutral biexciton 2X$^0$ (2 electrons + 2 holes). This attribution is based on measurements of PL intensity as a function of Laser excitation power (not shown) and analysis of the fine structure (Fig. \ref{fig:fig1}). In the absence of nuclear spin effects the X$^+$ shows as expected no fine structure \cite{belhadj09}. To verify if the conduction electrons recombine indeed with pure heavy hole states $\psi_{\pm 3/2}$, we measure the dependence of the linearly polarized PL intensity as a function of the angle with the crystallographic axis for the X$^+$, see Fig.\ref{fig:fig2} A.
In the X$^+$ ground state the two valence holes are in a spin singlet state (total spin S=0) and the anisotropic Coulomb exchange interaction with the conduction electron cancels out to zero. This makes the X$^+$ emission an ideal probe for the quantum dot symmetry \cite{koudinov04}. For a conduction electron recombining with pure heavy hole states the polar diagram in Fig. \ref{fig:fig2}A would show a perfect circle. We make two important observations: (i) a clear distortion of the circular pattern resulting in elliptical polarization (ii) neither the maximum nor the minimum of the ellipse are aligned with $[110]$ and $[1\bar{1}0]$ (note the tilt by an angle $\xi\approx 6^\circ$ in the figure).

There is no HH-LH mixing for the valence ground state of an unstrained, flat
cylindrical dot with circular base and perfectly symmetrical interfaces i.e.
the strength of the HH-LH coupling
$x=|I|/\Delta_{HL}\ll 1$ where the modulus of the matrix element $I$ couples HH and LH states in the Luttinger-Kohn (LK)
Hamiltonian.
To interpret our data, we have performed calculations based on only one deviation from this ideal dot: we assume an elliptical base with two symmetry axis in the $xy$ planes as suggested by AFM measurements. Along $z$ an infinite quantum well potential is used, in the $xy$ plane an harmonic oscillator potential \cite{warburton98}. For strain free dots only the LK Hamiltonian is taken into account, unlike for strained dots, where the Bir-Pikus Hamiltonian dominates \cite{koudinov04,leger07}. The resulting Hamiltonian is treated through the envelope
function approximation.
Concerning the elliptical polarization pattern, the recombination of electrons with LH states $\psi_{\pm1/2}$ is associated with photon polarization exactly opposite to that of HH transitions $\psi_{\pm3/2}$ (dotted arrows in Fig. 1A), resulting in elliptical polarization.
The mixed states are
$\tilde{\psi}_{\pm3/2}=\left(\psi_{\pm3/2}+\beta \psi_{\mp1/2}\right)/\sqrt{1+|\beta|^2}$
where $\beta = e^{i\xi}\left(1-\sqrt{1+4x^2}\right)/2x$. Subsequently the optical
selection rules with the usual oscillator strength ratio $f_{HH}/f_{LH}=3:1$ are applied.
The tilting of the polar diagram of the trion is controlled by $\xi=\mathrm{arg}\left(I\right)$.
The PL intensity as a function
of the angle of the analyzer in Fig.\ref{fig:fig2} A can be very well fitted by the function
\begin{equation}\label{noprime}
L\left(\theta\right)=c\left[a^2+b^2-2 a b\cos 2\left(\theta-\xi\right)\right]
\end{equation}
where
$a=\sqrt{x\left\vert\beta\right\vert/\left(3-6x\left\vert\beta\right\vert\right)}$,
$b=\sqrt{x/\left\vert \beta\right\vert \left(1-2x\left\vert \beta\right\vert\right)}$,
and the ratio between the electron-LH and electron-HH overlap integrals
$\left\langle\ \psi_{hh} \left\vert\right. \psi_e\right\rangle
/\left\langle\ \psi_{lh} \left\vert\right. \psi_e\right\rangle\approx 1$.
The fitting parameter $c$
is proportional to the oscillator strength of the
optical transition between pure HH and conduction states.
The ratio $\eta$ between the two axis of the elliptical polarization plot is
directly related to the HH-LH mixing coefficient and can be expressed as
$\eta=\left(\sqrt{3}-\left\vert\beta\right\vert\right)^2/
\left(\sqrt{3}+\left\vert\beta\right\vert\right)^2$.
The ratio $\eta$ being different from unity is a direct consequence of the elliptical base of
the quantum dot, see Fig. \ref{fig:fig2}A.
The fit of $L\left(\theta\right)$ with the experimental data shown in Fig. \ref{fig:fig2}A yields
$c=0.18$, $x=0.27$ and a tilting angle of $\xi\approx 6^\circ$.
This allows us to extract for the dot in Fig. \ref{fig:fig2}A $\eta \approx 0.57$ and a
strong mixing $|\beta|=0.25$ which lies within the typical range $0.16<|\beta|<0.3$ for
strain free dots in our samples, compared to $0.2<|\beta|<0.7$
reported for strained InAs, CdSe and CdTe dots \cite{koudinov04,Krizhanovskii,leger07}.

Concerning the observed tilting angle $\xi$ of the polarization plot with respect to $[110]$: this depends on the tilting angle $\phi$ of the main dot axis with respect to the crystallographic axis and on $\left(\gamma_2-\gamma_3\right)/\left(\gamma_2+\gamma_3\right)$
(which is taken to be is zero in the spherical approximation), where $\gamma_{2,3}$ are the Luttinger parameters.
As a consequence, the direction of the polarization tilt is not necessarily parallel to that
of the dot elongation.
The tilting angle $\xi$ of the polarization plot is zero if the quantum dot base is of perfectly circular shape and more interestingly, if $\phi=n\pi/4$, where $n \in \mathbf{Z}$.
Moreover it is maximal for $\phi=\pi/8+n\pi/4$.
This is a direct consequence of the four-fold rotational symmetry of the crystal structure.

As expected the X$^0$ shows a fine structure, see Fig. \ref{fig:fig1}. The two bright X$^0$ states
$ \vert X\rangle = \left(\vert \Uparrow, \downarrow \rangle
 +\vert \Downarrow, \uparrow \rangle\right)/\sqrt{2}$
and
$\vert Y\rangle = \left(\vert \Uparrow, \downarrow \rangle
 -\vert \Downarrow, \uparrow \rangle\right)/i\sqrt{2}$
are separated in energy by $\delta_1 \equiv E_X - E_Y$ due to anisotropic electron hole
exchange \cite{bayer02}.
Here  $\Uparrow (\Downarrow)$ stands for the heavy hole pseudo spin up (down)  and $\uparrow (\downarrow)$ for
the electron spin up (down) projections onto the $z$-axis. The same splitting $\vert\delta_1 \vert$ is found for the 2X$^0$, but as expected with the order of the peaks reversed.
The results for the X$^0$ are due to competition between structural asymmetry of the dot and asymmetry of the electron hole exchange. For pure heavy hole states, two states $\vert Y\rangle$ and $\vert X\rangle$ are polarized along $[110]$ and $[1\bar{1}0]$. Our diagram in Fig. \ref{fig:fig2}B shows that the actual eigenstates
of the system are given by
$ \vert X'\rangle=\cos\varphi\vert X\rangle-\sin\varphi\vert Y\rangle$
and
$ \vert Y'\rangle=\sin\varphi\vert X\rangle+\cos\varphi\vert Y \rangle$
where $2\varphi$ is the argument of the matrix element that couples
the two HH exciton states due to anisotropic exchange interaction.
The emission intensity of the $\left\vert X'\right\rangle$
and $\left\vert Y'\right\rangle$ excitons
as a function
of the angle of the analyzer can be expressed as
\begin{eqnarray}
L_{X'}\left(\theta\right)
&=&c\left[a \cos\left(\theta+\varphi-2\xi\right)
+b\cos\left(\theta-\varphi\right)\right]^2,\label{xprime}\\
L_{Y'}\left(\theta\right)
&=&c\left[a \sin\left(\theta+\varphi-2\xi\right)
-b\sin\left(\theta-\varphi\right)\right]^2\label{yprime}.
\end{eqnarray}
In Fig. \ref{fig:fig2}B the thick lines indicate
the polar diagrams of the
$\left\vert X'\right\rangle$ (gray)
and $\left\vert Y'\right\rangle$ (black)
excitons fitted with the same parameters as for the trion
in Fig. \ref{fig:fig2}A underlining the consistency of our model.
We observe two transitions $\vert Y'\rangle$ and $\vert X'\rangle$ that
are not aligned along $[110]$ and $[1\bar{1}0]$.
Although the angle between the polar diagrams
of $\vert Y'\rangle$ and $\vert X'\rangle$
in Fig. \ref{fig:fig2}B is approximately $90^\circ$, the
theory allows for non
perpendicular diagrams.

In summary we have measured the neutral and charged exciton emission
intensity of GaAs quantum dots grown by droplet epitaxy.
We studied the emission of both quasiparticles within the
LK Hamiltonian theory and the envelope function
approximation.
First we have shown that
the narrowing of the polar diagram of the charged exciton's emission
intensity depends strongly
on the shape asymmetry of the quantum dot that is responsible
for the mixing of HH and LH states (through the Luttinger-Kohn Hamiltonian
matrix element $I$) even in the absence of strain.
Interesting new effects could arise from
further anisotropy (electric field or composition gradient along the $z$ axis)
coming from coupling between bright and dark exciton states.
Second we have found that the tilting of the polar diagram
is not in general parallel to that of the dot elongation
and depends on
(i) the difference in the $\gamma_2$ and $\gamma_3$ Luttinger parameters
and (ii) the angle between the main axis of the quantum dot
and the $[110]$ direction of the crystal; under the spherical approximation,
i.e. $\gamma_2=\gamma_3$, the calculated tilting is zero.
Third, we obtained expressions for the PL polar intensity of the $\left\vert X'\right\rangle$
and $\left\vert Y'\right\rangle$ exciton states that are not in general perpendicular.
The angle between them is tuned by the anisotropic exchange interaction
between the hole and the electron.
Our work suggests that the strong effect of shape anisotropy
on HH-LH coupling should also be included in
a complete analysis of PL polarization
in strained Stransky-Krastanov type QDs.

We thank ANR-QUAMOS, IUF, DGA
for  support and Yoan L\'eger for stimulating discussions.
A. Kunold acknowledges financial support from
``Estancias sab\'aticas al extranjero'' CONACyT and ``Acuerdo 02/06'' Rector\'{\i}a UAM-A,
for his sabbatical stay from UAM-A in INSA Toulouse


\end{document}